\documentclass[lettersize,journal]{IEEEtran}
\usepackage{amsmath,amsfonts}
\usepackage{algorithmic}
\usepackage{algorithm}
\usepackage{array}
\usepackage[caption=false,font=normalsize,labelfont=sf,textfont=sf]{subfig}
\usepackage{textcomp}
\usepackage{stfloats}
\usepackage{url}
\usepackage{verbatim}
\usepackage{graphicx}
\usepackage{xcolor}
\usepackage{cite}
\usepackage{tabularx}
\begin{document}

\title{Convergence of Manufacturing and Networking\\in Future Factories}
\author{Ilaria Malanchini, Nicola Michailow, Patrick Agostini, Janne Ali-Tolppa, David Hock, Martin Kasparick, Alessandro Lieto, Nikolaj Marchenko, Alberto Martinez Alba, Rastin Pries, Qiuheng Zhou
\thanks{Ilaria Malanchini, Janne Ali-Tolppa, Alessandro Lieto, and Rastin Pries are with Nokia Solutions and Networks. Nicola Michailow and Alberto Martinez Alba are with Siemens Technology. Patrick Agostini, and Martin Kasparick are with Fraunhofer Heinrich Hertz Institute. David Hoch is with Infosim. Nikolaj Marchenko is with Robert Bosch GmbH. Qiuheng Zhou is with the German Research Center for Artificial Intelligence.}
}

\markboth{}%
{Shell \MakeLowercase{\textit{et al.}}: A Sample Article Using IEEEtran.cls for IEEE Journals}


\maketitle


\begin{abstract}
The roll out of 5G has been mainly characterized by its distinct support for vertical industries, especially manufacturing. Leveraging synergies among these two worlds, namely production facilities and network systems, is a fundamental aspect to enable flexibility and economic viability in future factories. This work highlights the potential for intelligent networking and advanced machine learning-based solutions in 5G-and-beyond systems in the context of Industry 4.0 and flexible manufacturing. The intersection thereof allows to create versatile machines and dynamic communication networks that can adapt to changes in the manufacturing process, factory layout and communication environment, supporting real-time interaction between humans, machines, and systems. We present a vision and corresponding framework by introducing the network-aware and production-aware principles, outlining results achieved in this context and summarizing them into three key use cases. Finally, we discuss a selection of remaining open challenges in private networks as well as give an outlook on future 6G research directions.
\end{abstract}


\section{Introduction}\label{sec:1}
Optimal capacity utilization and reduced factory downtimes are highly desirable goals in manufacturing. High expectations are placed on the paradigm of flexible manufacturing~\cite{IECref}, i.e., the capabilities to produce highly customizable products and small batch sizes, to ensure economic viability in upcoming years. It is envisioned that future factories will be able to reconfigure based on manufacturing needs within short periods of time (e.g., few days). In addition, autonomous cyber-physical systems (including mobile robots), will be instrumental in shaping the future of manufacturing by enhancing flexibility, productivity, safety, real-time monitoring, and collaborative capabilities within factory environments. This transformation will lead to more efficient, adaptable, and responsive production processes, ultimately driving innovation and growth in various industries.
Two important ingredients for achieving such goal are versatile machines that require little or no human supervision and dynamic communication networks that can adapt to deliver services even when the manufacturing process or the layout of the factory changes. Additionally, the production and the communications domains must tightly interact and support each other.

5G has been a key driver of the digitalization in Industry~4.0 and use cases from vertical industries~\cite{5GACIA}, such as manufacturing, contributed to the definition of 5G performance targets, including latency and reliability. One novelty of 5G is the paradigm of non-public networks, i.e., cellular networks that cover limited geographic areas using a dedicated frequency band. Operating such private networks is expected to offer connectivity with predictable performance along with advantages in terms of reliability and security. Such networks may be deployed, e.g., in factories, at construction sites or on agricultural land.

Machine learning (ML) algorithms are trained instead of being explicitly programmed and are able to learn solutions to problems that are too complex for human experts or classical optimization methods. These are highly desirable properties in a world of ever-increasing technological complexity, and in particular towards reconfigurable factories and the adaptive communication networks therein. The trade-off is that ML approaches heavily depend on the availability and quality of domain- and application-specific training data.
Factories present a unique communication environment that favors the use of machine learning. Usually, there is a controlled setup with deterministic or well predictable patterns that occur as a consequence of periodic and isochronous traffic. Furthermore, there is a strong correlation between the communication patterns and the factory and production processes. Challenging objectives in such highly dynamic industrial environments are to guarantee reliable wireless connectivity and edge cloud integration, to facilitate flexible reconfiguration of communication networks, to adapt to the dynamicity of heterogeneous network services and to account for variations that arise from moving parts, goods, and robots.

\begin{figure}[t]
\includegraphics[width=\columnwidth]{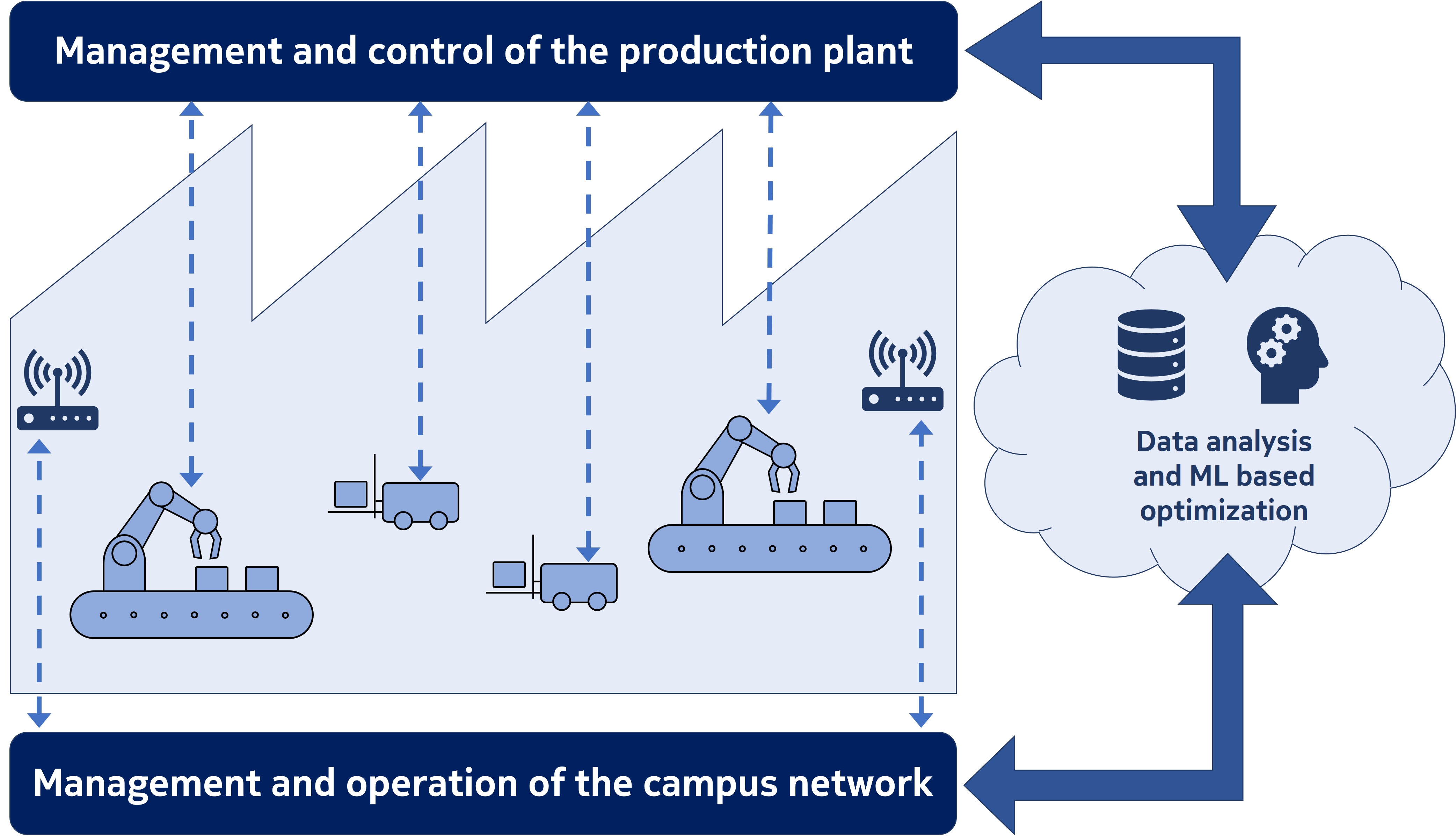}
\centering
\caption{Data analysis and ML based optimization serve as enablers to leverage synergies between the production and the communication systems.}
 \label{Fig:1}
\end{figure}

This work highlights the potential for intelligent networking and advanced machine learning-based solutions in 5G-and-beyond systems in the context of Industry 4.0 and flexible manufacturing. It presents methods to harvest synergies between industrial applications and industrial communications through ML, which is a suitable tool to detect and exploit hidden patterns and to find non-trivial correlations across these two domains (as depicted in Fig.~\ref{Fig:1}). We elaborate on specific examples of how 5G and ML technologies can be combined to develop solutions that tackle: 
\begin{itemize}
    \item Network-aware production optimization, i.e., how to predict and exploit key performance indicators (KPIs) in networks to optimize production applications and services.
    \item Production-aware network optimization, i.e., how to obtain and incorporate production status/information into the network management system to optimize mechanisms such as network slicing.
\end{itemize}

The following sections elaborate on these two perspectives and present three case studies: i) Prediction of network KPIs to optimize production system efficiency and in particular edge cloud offloading, ii) Production-aware resource allocation for slicing and iii) Radio map, localization, and end-to-end monitoring for intralogistics in flexible manufacturing. Those are followed by a discussion on open challenges in private networks and an outlook on possible evolution towards 6G.

\section{Production and Network Convergence}\label{sec:2}
Differently from the the integration of information technology (IT) systems with operational technology (OT) systems, the so-called IT/OT convergence, which has been investigated and implemented since several years, the convergence of the manufacturing (i.e., OT) domain and the network domain is still in its infancy. However, it is essential to exploit synergies of the two worlds to realize the full potential of automated production that is expected in factories of the future. 
Hereafter, we elaborate on how ML can enable such convergence, and, based on the vision and framework developed within the project KICK (``\textit{Künstliche Intelligenz für Campus Kommunikation}'')~\cite{KICK}, we detail the two directions of such interaction (namely, network-aware production and production-aware network).

First, a fully integrated network-OT system has to rely on ML-based principles to collect and exploit data coming from different sources, i.e. different domains. Therefore, such system has to enable the following steps:
\begin{itemize}
    \item \textit{Data acquisition}: Data is collected from various sources coming from both the production and the network domain, this is then stored and managed possibly by a central database and fed into the learning algorithms. To deal with scarcity of data in real environments, one possibility is to enrich the data set with artificially generated data and generative data models.
    \item \textit{Modeling}: When looking at the industrial environment, synergies among production and network can be exploited in two directions: by using the production information to optimize the network or by using the network information to optimize the production processes. Modeling a specific problem includes also identifying how such synergies can be leveraged in either one or the other direction, as well as which (and how) information should be exchanged.
    \item \textit{Solving}: For any given production/network model, the most appropriate solving approach needs to be identified. ML is a powerful tool to create a learning environment and to enable a continuous exchange of information and feedback between the two domains. In this regard, deep reinforcement learning and transfer learning are two of the most valuable assets, especially to exploit cross-domain correlations as well as to adapt to the continuous changes of the factory environment.
    \item \textit{Testing}: The proposed algorithms are usually tested at first in a simulation-based environment reproducing  the production processes and facilities, the so-called digital twin. Specific joint network/production simulators/digital twins are needed to test the reciprocal influence between the two domains. Once the algorithms are tested in a simulation environment, the transfer to a realistic testbed is desirable. 
    \item \textit{Closing the loop}: Data collected during testing (either generated from a simulator or from the real environment) as well as feedback on the network and production performance can and should be used to fine tune the proposed ML models and corresponding algorithms. This continuous feedback closes the loop and creates the so called ``closed-loop automation.''
\end{itemize}

\subsection{Network-aware Production Optimization}
With network-aware production optimization we refer to the use of information about the (present and future) state of the communication networks to optimize production processes. Providing reliable wireless connectivity is key to many industrial applications. Current pre-planned networks with fixed topologies and static configurations might not be able to satisfy the requirements of future applications. In fact, a precondition for the flexibility of future industries is the provisioning of ultra-high performance radio links, which allows to cut loose machines from the restraints of interconnection cables. ML and ML-based digital twins enable the accumulation of data from both the wireless networks and the manufacturing processes. 
This knowledge can be shared among machines and robots to help optimize the dynamic and evolving manufacturing processes. For instance, once could use information on possible degradation of the quality of service (QoS) to avoid sudden session interruption of critical applications (e.g., safety, automated driving). %
However, wireless networks exhibit variations in QoS over space and time, and the variance can be a function of observed or unobserved quantities. In stationary environments, such variability can be adequately characterized by second order (channel) statistics. 
Therefore, since many industrial applications have strict requirements on the quality of service provided by the communication networks, the actual execution of the application can be adapted by taken into account both the predicted communication QoS as well as corresponding second order statistics. Furthermore, to support dynamic applications, ML can also be used to predict time instances or locations with favorable communication properties. Those can be leveraged to shape the data transmission pattern of industrial applications while improving their reliability. An example is edge cloud offloading, which is described in detail further below, where certain tasks can be offloaded to an edge cloud or performed locally, depending on the (predicted) network state.

\subsection{Production-aware Network Optimization}
Similar to the wireless network performance, also the requirements of industrial applications vary in space and time, depending on the state of the underlying production processes. Translating such production requirements into constraints on the communication network is a challenging task. Nevertheless, adapting the network to the production needs is crucial since static resource allocations and network configurations may lead to over- or under-provisioning and may not be able to accommodate changes in the factory environment. Introducing information on production states into network optimization routines in general, and the provisioning of communication resources in particular, allow the network to fit the applications’ instantaneous requirements, thereby increasing system efficiency. Moreover, introducing performance and safety metrics from the production as constraints into the network management leads to increased service reliability and enables safer operations of industrial applications. An example of production-aware network optimization is described in the following section, where network slicing resources are allocated based on production state and requirements.

\section{Key Use Cases}\label{sec:3}
In this section, we present three key use cases that show how the interworking between production and network systems can lead to significant gains on both sides.

\subsection{Network Prediction for Production System Efficiency}
Energy efficiency in the industrial world is a key aspect not only for meeting the needs of a sustainable environment, but also for increasing profitability. The availability of high-performance computing platforms, combined with the ability to run artificial intelligence algorithms in real time, has sparked a continuous transition to move most of the processing towards the edge (i.e. edge computing). Moving the processing from the devices to the edge does not only provide higher processing power, but also reduces the energy consumption of devices thereby increasing battery life. Furthermore, it frees the limited computational resources onboard the devices for other tasks. 
One example is the edge cloud offloading of vision-based positioning (VBP) algorithms for automated guided vehicles (AGVs). Offloading the simultaneous localization and mapping (SLAM) VBP algorithm from the AGV to the edge cloud saves onboard computing resources and energy, reduces the required computation time, and enables combining maps from several devices, leading to a higher positioning accuracy~\cite{Marchenko2022}.

One of the main aspects to fully exploit the benefits of edge cloud offloading is finding the right compromise between processing performed on the edge and local processing. Indeed, the offloading task requires sufficient uplink (UL) throughput and low-enough network latency not to introduce unacceptable delay to the application. As the achievable throughput is subject to spatio-temporal fluctuations, offloading attempts can fail, causing higher latency, increased communication overhead and risk of harming the production processes. To mitigate offloading failures, offloading decisions must be robust against the variability of the achievable network QoS and corresponding KPIs (such as UL throughput). Furthermore, the onboarding and offloading operations typically require certain amount of time to complete, which also needs to be taken into account in the edge cloud offloading orchestration. Learned spatio-temporal QoS fluctuations can be used to predict optimal onboarding or offloading points, where both traditional algorithms, e.g., k-nearest neighbors (k-NN) and random forest, as well as more advanced deep learning-based approaches, e.g., convolutional neural networks (CNNs), have been shown to achieve competitive performance, provided that the distribution of the UL throughput remains stationary, over space and time. A comparison among those different options has been derived considering a setup as the one described in~\cite{Marchenko2022} by using real data collected in a factory. Results  in Table~\ref{table1} show that by introducing prediction of the QoS in the decision process we can achieve considerable gains with respect to the reactive case (where no KPIs prediction is used). In particular, CNN provides the highest offloading time and the least number of replacements, whereas all three predictive approaches provide similar performance in terms of number of offloading failures. 

\begin{table}[t]
\caption{Performance Comparison for Edge Cloud Offloading Task}
\label{table1}
\begin{center}
  \begin{tabular}{ | c | c | c | c |}
    \hline
    \textbf{Algorithm} & Offloading & Re-placements & Failures \\ \hline\hline
    Oracle & 81.8 & 155 & 0\\ \hline
    Reactive & 50.18 & 254 & 21 \\ \hline
    12-NN Regression & 59.01 & 162 & \textbf{6} \\ \hline
    Random Forest & 58.55 & 137 & 7 \\ \hline
    CNN & \textbf{60.4} & \textbf{135} & 7 \\ \hline
  \end{tabular}
\end{center}
\end{table}

Offloading decisions must be robust also against prediction uncertainties by accessing the probability of false-positives to guarantee stable localization performance and minimize communication overhead. 
Here, probabilistic approaches such as Gaussian processes or mixture density networks, provide powerful learning capabilities for assessing aleatoric and epistemic uncertainties, which can be leveraged in intelligent decision-making processes.
Accurate and up-to-date channel state information (CSI) is a fundamental requirement which powers wireless adaptation strategies that enable reliable and near Shannon limit communication. Especially in industrial scenarios which involve high mobility, i.e., AGV communication, CSI is subject to high variability, which increases the need for frequent CSI updates, incurring a non-negligible system overhead. In stationary environments, machine learning driven CSI prediction can be leveraged to reduce the required update frequency by observing that in many scenarios, CSI is characterized by high spatial correlation. To this end, ML methods for time-series prediction provide suitable learning algorithms from which especially deep learning-based approaches have been shown to provide competitive performance. By using a deep learning architecture, which consists of a one-dimensional CNN layer, a long-short term memory (LSTM) layer, and a dense output layer, highly accurate CSI predictions can be achieved. By exploring the prediction performance on datasets from different propagation environments collected from an end-to-end 5G new radio communication platform, we observe that in relatively open indoor environments, the achieved prediction performance can help reducing the frequency of channel reports\cite{Zhou2022}. However, in a closed indoor environment, the coherence time of CSI is drastically reduced due to multi-path propagation effects, which reduces prediction performance and, consequently, potential savings.

In practice, network KPIs distribution may not be stationary on a large time scale and change over time. Detecting shift of distributions and adapting learned models appropriately is an open challenge which needs further considerations from methodological and practical point of view. KPIs probing in combination with probabilistic models can enable assessing distributional shifts in a statistical sense. Furthermore, approaches from continual learning may be leveraged to adapt learned models to the new distribution in an online and data efficient manner.

\subsection{Production-aware Resource Allocation for Slicing}
Network slicing is a practical way to handle different types of services/applications on the same network while making sure each one of those can meet specific requirements and fulfill QoS expectations. Although network slicing has been initially considered to isolate the network infrastructure of public networks in separated and independent slices, it may have an important role also in industrial private networks, by giving the possibility to isolate the traffic belonging to different production processes, contributing to the monitoring and optimization within the factory. Particularly useful is the case where resources are shared dynamically among slices, since it allows to use the resources at their best. However, especially in industrial environments it is essential to meet the stringent requirements of the most demanding and critical applications, even when performing dynamic slicing. 

To dynamically tailor the network slice configuration and resource allocation to different industrial applications and corresponding requirements, we propose to exploit the repetitive traffic patterns and  introduce the so-called ``production status'' of the industrial environment. By production status we refer to a representation of the factory floor and current operational modes of the machines and industrial devices, e.g., which machines are active, which AGVs are moving, current production tasks, etc. The repetitiveness of the industrial operations and, consequently, of the production status create cyclic traffic patterns that are highly predictable and can therefore help the system to learn the optimal slice resource allocation policy. 

\begin{figure}[t]
\includegraphics[width=0.8\columnwidth]{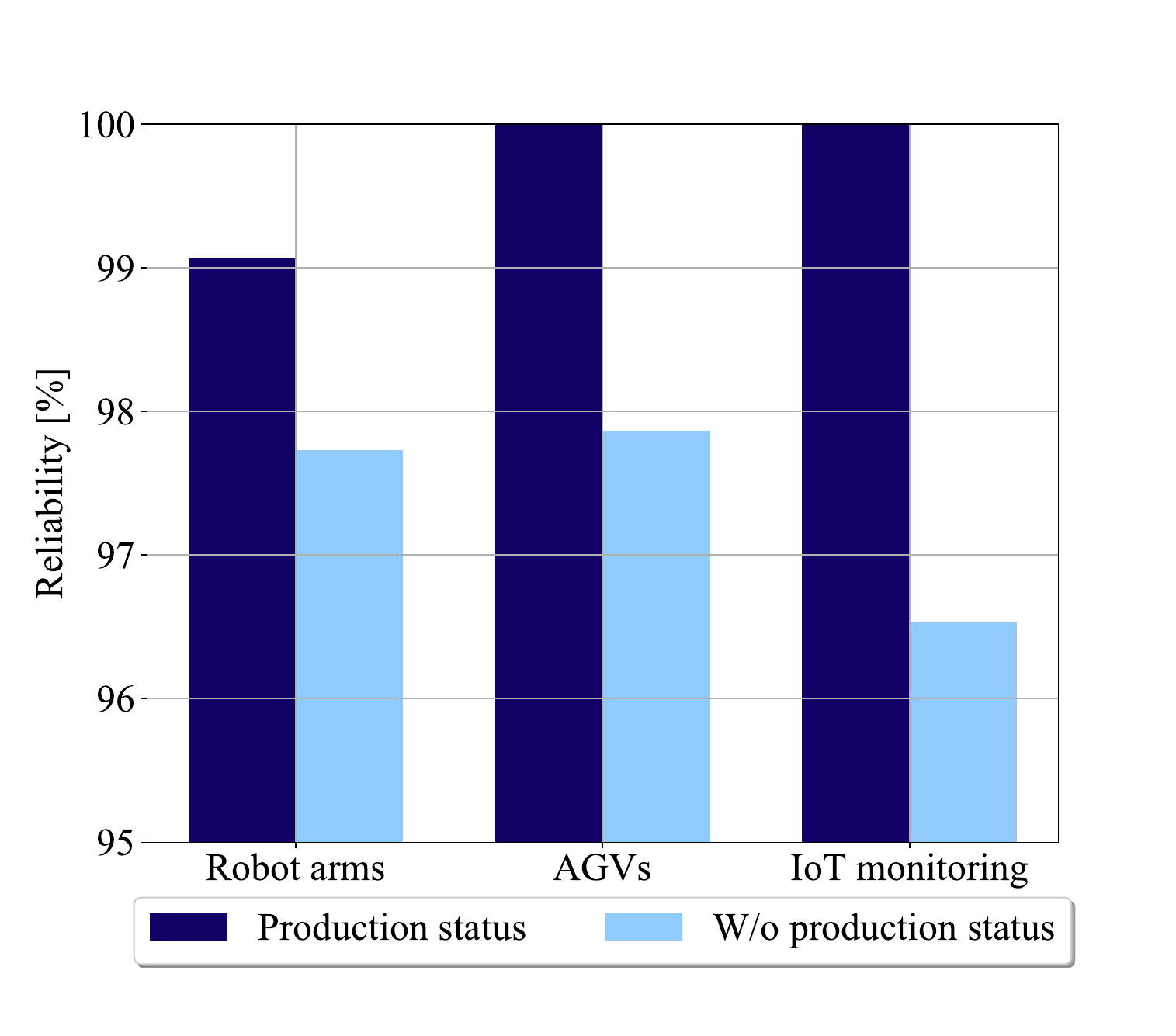}
\centering
\caption{Reliability of slices with and without production status information.}
 \label{Fig:ReliabilitySlices}
\end{figure}

In particular, we propose a reinforcement learning (RL) agent that, by merging data collected from network and production, learns how to dynamically allocate resources to different slices to meet the industrial requirements and, at the same time, maximizes resource efficiency. 
Namely, we assume that the agent can collect observations from the network domain, e.g., propagation channel conditions, achieved throughput and latency, packet loss rate, together with observations from the industrial environment, e.g., number of active devices dedicated to a specific task and their operational status. By introducing production status into the learning model, the RL agent is able to predict the variations in traffic demand or channel propagation conditions induced by the production dynamics and adapt the resource allocation accordingly, therefore providing better performance, in terms of higher resource efficiency, reduced latency and higher reliability. Fig~\ref{Fig:ReliabilitySlices} shows the improvement in terms of reliability (i.e. probability of meeting latency and throughput requirements) of three different slices when comparing the performance of the RL agent with and without production status information, using a setup as described in~\cite{Zambianco2022}. 
Additional types of production status information can be directly derived from performance metrics of industrial applications, which oftentimes are readily available, i.e., for closed loop applications, or can be learned with ML techniques. The availability of such metrics can be leveraged in learning constraint network slicing policies which adhere to the performance metrics of industrial applications and provide safe operation mode of application at the same time~\cite{Agostini2022}.

Although the proposed RL approach works well on repetitive scenarios with short time scale dynamics and can be useful for online implementation of resource allocation algorithms, it might fall short in capturing macro, longtime scale changes of the factory. One of the requirements of future factories is the possibility of, e.g., changing the production line of a specific plant over a weekend. This implies changes in user positions and operations, factory layout and, therefore, changes in the radio propagation maps as well as in the traffic patterns and, in some cases, even performance requirements. Transfer learning (TL) can be employed to enable fast re-adaptation of the slicing policies to the new production setup. As a matter of fact, TL techniques can transfer the knowledge acquired from a given environment into a new environment, by exploiting the similarities in the data distribution of the two domains. An example of resource adaptation for network slices to a new environment has been studied and proposed in~\cite{Hu2022}, where the pre-trained model used to learn the resource allocation policy of a set of base stations has been transferred to learn the resource allocation policy for a different set of base stations, where the base stations differ in size, coverage area and traffic dynamics. Results show that transfer learning allows for a better initial starting point (especially useful for online learning), faster convergence – approximately needing $33\%$ less epochs to converge – and better network performance. 

These results are just an initial step towards leveraging the tremendous potential of joint production and network optimization. We envision that even bigger gains could be achieved when providing continuous feedback and setting a common goal between the two systems. 

\subsection{Intralogistics for Flexible Manufacturing}
Consider a flexible manufacturing scenario, where fixed production lines and predetermined flows of goods and products are replaced by reconfigurable manufacturing stations, autonomous vehicles, and self-guided products\cite{AIMFREE}. In such environment, potentially for every order, the path of the goods, the function of the machines, or the layout of the factory may need to change. In many cases, wires would inhibit the required flexibility, hence this is a prime use case for wireless access technologies like private industrial 5G to complement fixed communication infrastructure in a factory. 

To guarantee an adequate connection quality between access points and mobile receivers such as AGVs, it is useful to predict the signal strength received from each radio transmitter at every point in a considered area, e.g., shop floor. In other words, we desire a radio environment map (REM) to be estimated for each transmitter within the factory, which should be updated in real time to reflect dynamic changes in the physical environment. Such REM can assist human personnel with (manual) planning of network deployments, wireless nodes with making resource allocation decisions and mobile devices with reacting to disruptions to their connectivity. 

There are two sources of information that we can exploit to generate such REM. 
First, since the layout, machines, and materials within the factory are known, we can employ theoretical propagation models to find out the expected received signal power at each point. There are multiple such models, ranging from simple distance-based models to more advanced approaches such as the Motley-Keenan model or ray-tracing simulations\cite{Calle2012}. Advanced models offer more accurate predictions, but the corresponding computational time may be exceedingly large for factory areas, especially if the influence of moving AGVs or other parts on the radio map must be considered. To this end, it is possible to use machine learning techniques to produce sophisticated radio maps as those generated by ray-tracing simulations at a fraction of the computing time and resources. For instance, recent research shows that this can be accomplished by training U-Net convolutional neural networks to generate outdoor and indoor radio maps~\cite{Levie2021}. 
The second source of information for generating accurate indoor radio maps are dedicated measurements. Instead of relying on theoretical models, a set of reference receivers can be deployed within the factory to sample the signal strength for every transmitter. Then, machine learning approaches can be used to reconstruct a model-free radio map, leading to more efficient computation and better results than conventional approaches. Finally, it is possible to fuse both model-based and model-free approaches to yield a high-quality radio map, which can be employed to optimize the behavior of the AGVs. 

\begin{figure}[t]
\includegraphics[width=\columnwidth]{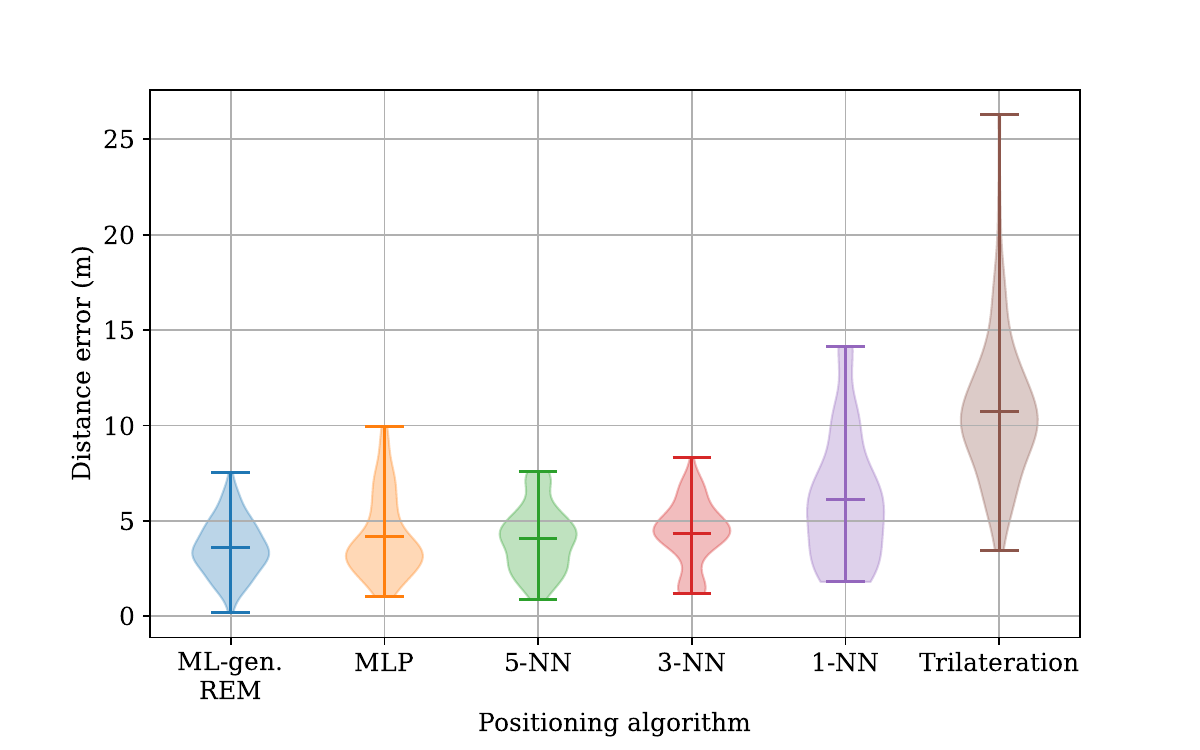}
\centering
\caption{Distance error for different indoor positioning algorithms via signal strength fingerprinting.}
 \label{Fig:uncertainty}
\end{figure}

In addition, the existence of accurate up-to-date REMs from multiple transmitters can be used to pinpoint the location of an AGV via signal strength fingerprinting. That is, given enough transmitters distributed over the factory, each point in the considered area can be identified by a set of statistically different signal strength measurements, one per transmitter. Recent work shows that ML approaches can result in better positioning accuracy than conventional approaches. By adopting a setup as the one described in~\cite{Martinez2022}, the distance error behavior is shown in Fig.~\ref{Fig:uncertainty}, where the ML-generated REM is compared to a multilayer perceptron (MLP), to the k-NN algorithm, for different values of k, and to trilateration. Accurate positioning of moving AGVs within the factory is crucial for efficient and flexible manufacturing since it enables fast and reliable reconfigurations.

Another essential feature in flexible manufacturing is accurate end-to-end (E2E) monitoring, to guarantee that the network behavior meets the expectations of the factory owner even when the underlying physical layout changes. A straightforward but cumbersome approach to this end is to add monitoring agents to all switches and devices in the network. These agents collect network statistics that can be used to describe the overall network behavior and identify anomalies. A smarter, more efficient approach is to use ML techniques to leverage one crucial observation: the traffic behavior in one network region usually provides information about the state in other regions. As a result, an ML algorithm can learn the relationships among network regions and thus, it can be used to estimate the state of regions with few monitoring agents. A possible strategy to accomplish this is to model time-dependent traffic patterns as Gramian angular fields, in order to be effectively handled by image-processing ML algorithms. This overall approach combines the production-aware network perspective, since knowledge from the network characteristics are used to enhance network monitoring, and the network-aware production optimization perspective, as E2E monitoring improves production efficiency and resilience.

In the bigger picture, both REM generation and E2E monitoring contribute to the creation of a digital network twin that collects current measurements and estimations of the state of the physical world (i.e. factory environment) and stores a history of past states, paving the way for future applications of ML in this area.

\subsection{Challenges in Private Networks}
From the experience gained during our project\cite{KICK}, we have been able to identify and partially address the following challenges related to production and network convergence in private networks for manufacturing.
\begin{enumerate}
    \item \textit{Joint production and network data collection}: To model and test network-aware and production-aware optimization algorithms, data sets comprising both production and network data are of vital importance. We developed a production/network simulator able to re-create an industrial environment, simulate several production processes and replicate an indoor 5G-based wireless network. Thanks to the simulator, joined production and network data can be generated and used for training the proposed ML algorithms~\cite{Lieto2022EW}. To support further developments in this direction, the KICK simulator has been made available to the research community\footnote{The KICK simulator can be downloaded at: https://gitlab.hhi.fraunhofer.de/simulator/kick-process-simulator}.
    \item \textit{Traffic models in industrial environments}: To enable realistic testing of the developed algorithms, simulation environments should use models that are as close as possible to reality. One challenge in industrial environment is the lack of real data and corresponding realistic traffic models. To overcome this, we collected measurements from a factory from Trumpf and proposed generative models to be used in simulating environments\cite{Lieto2022}. The derived generative models have also been made publicly available and can be used to reproduce realistic industrial traffic\footnote{The industrial traffic models derived from a real factory can be downloaded at: https://github.com/nokia/genIndustrialNetTraffic}.
    \item \textit{Testing in a real factory environment}: Factory environments are characterized by a lot of cluttered, highly reflective surfaces (as metallic pipes, machines) as well as highly absorbing surfaces (as concrete). Also, they are affected by short-term changes due to, e.g., replacements, storage, maintenance. Making wireless network tests in a real factory environment can help to obtain realistic radio and performance measurements that can be used for the data-driven model development and system simulation. Within the context of our project, we build a 5G testbed in a real Bosch facility and experimented different use cases, such as KPIs prediction, ray-tracing model calibration and indoor network planning~\cite{Liao23GL}. 
\end{enumerate}

Nevertheless, it remains still a big challenge to tightly integrate the wireless network operations and production processes, due to the intrinsic requirement for an extremely reliable and continuous interworking. It is desirable, in the future, to have hybrid approaches where real data from the production is used to fine-tune network emulators and factory digital twins such that continuous feedback is provided between communications and manufacturing systems and network optimization is performed taking into account production goals.

\section{Conclusion}\label{sec:4}
By bringing together manufacturing infrastructure and communication networks, the topics discussed in this paper are only the first step towards the path that will lead to emerging 6G technologies for future factories. In future communications and production systems, the interaction of heterogeneous entities and embodied ML agents  in the shape of autonomous collaborating cyber-physical systems is expected to give rise to conglomerates of subnetworks and networks of networks. Even more complex management and orchestration systems will be needed to manage the coordination among several domains and stakeholders. Contributions of 6G technology to the convergence of communications and sensing in future wireless networks is anticipated, allowing a tight integration of systems, in particular in manufacturing and network environments. In this context, advanced real-time digital twins will likely play a major role, e.g., for real-time network and production optimization, as well as for extended reality applications. The approach of sub-THz frequencies will pave the way to higher data rates, lower latencies, more devices per area, and enhanced sensing and localization capabilities, while the concept of semantics will change how and what information is exchanged among future autonomous robotic systems.


\section*{Acknowledgments}
This work has been supported by the German Federal Ministry of Education and Research (BMBF) project KICK.

\bibliographystyle{IEEEtran}
\bibliography{bibliography}

\vfill

\end{document}